\documentclass[submission, PhysProc]{SciPost}

\hypersetup{
    colorlinks,
    linkcolor={red!50!black},
    citecolor={blue!50!black},
    urlcolor={blue!80!black}
}

\usepackage[bitstream-charter]{mathdesign}
\usepackage{subcaption}
\usepackage{caption}

\DeclareSymbolFont{usualmathcal}{OMS}{cmsy}{m}{n}
\DeclareSymbolFontAlphabet{\mathcal}{usualmathcal}

\begin{document}

\begin{center}{\Large \textbf{
Forward-backward correlations in proton-proton collisions at the LHC energy: A model based study\\
}}\end{center}

\begin{center}
Joyati Mondal\textsuperscript{1$\star$},
Somnath Kar\textsuperscript{1$\star$}, 
Hirak Koley\textsuperscript{1},
Srijita Mukherjee\textsuperscript{1}, 
Argha Deb\textsuperscript{1,2} and 
Mitali Mondal\textsuperscript{1,2 $\star$}
\end{center}

\begin{center}
{\bf 1} Nuclear and Particle Physics Research Centre, Jadavpur University, Kolkata 700032, India
\\
{\bf 2} School of Studies in Environmental Radiation and Archaeological Sciences, Jadavpur University, Kolkata 700032, India
\\
* joyati254@gmail.com
* somnathkar11@gmail.com
* mitalimon@gmail.com
\end{center}

\begin{center}
\today
\end{center}


\definecolor{palegray}{gray}{0.95}
\begin{center}
\colorbox{palegray}{
  \begin{tabular}{rr}
  \begin{minipage}{0.1\textwidth}
    \includegraphics[width=13mm]{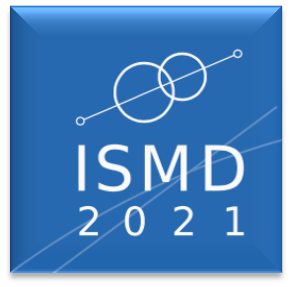}
  \end{minipage}
  &
  \begin{minipage}{0.75\textwidth}
    \begin{center}
    {\it 50th International Symposium on Multiparticle Dynamics} {\it (ISMD2021)}\\
    {\it 12-16 July 2021} \\
    \end{center}
  \end{minipage}
\end{tabular}
}
\end{center}

\section*{Abstract}
{\bf
Forward-backward (FB) multiplicity and momentum correlations of produced particles between symmetrically located pseudorapidity ($\eta$) intervals have been studied using the $p$QCD inspired EPOS3 model with and without hydrodynamical evolution of particles in proton-proton ($pp$) collisions at the center-of-mass energy, $\sqrt{s}=13$ TeV. The pseudorapidity-gap ($\eta_{gap}$) dependence of FB correlation strength is compared with our previously published results at $\sqrt{s}=$ 0.9, 2.76 and 7 TeV. The study reveals that the general trends of FB correlation strength at $\sqrt{s}=13$ TeV are similar to our previous observations at lower center-of-mass energies. We also find that the $\delta\eta$-weightage average of FB correlation strength as a function of different center-of-mass energies ($\sqrt{s}=$ 0.9, 2.76, 7 and 13 TeV) tends to saturate at very high energy.
}
%
\section{Introduction}
\label{sec:intro}
Forward-backward correlation, a robust tool, plays important role in understanding the dynamics of multiparticle interactions and their hadronization in different collision systems. Various experiments including CERN Super Proton Synchrotron (SPS) \& the Large Hadron Collider (LHC), the Tevatron, the Relativistic Heavy Ion Collider (RHIC) observed sizeable FB correlation strength in heavy-ion collisions as well as in small collision systems.\\ In this contribution, we have studied the FB multiplicity and momentum correlation in $pp$ collisions at $\sqrt{s}=13$ TeV using the Monte-Carlo event generator, EPOS3 \cite{ref1, ref2} featuring with and without hydrodynamical evolution (abbreviated as ``with and without hydro" in the rest of the texts) of particles. We have compared our results with our previously published results using the same model at $\sqrt{s}=$ 0.9, 2.76 and 7 TeV~\cite{ref3} and with the available Quark-Gluon String Model (QGSM) predictions at the top LHC energy~\cite{ref4}.
\section{Forward Backward Charged Particle Correlation Coefficient}
\label{sec:fbcorrCoeff}
Generally, FB correlations can be categorized into three main types \cite{ref5}: correlation between charged-particle multiplicities $(n - n)$, between mean or summed transverse momenta of charged particles $(p_{T} - p_{T})$ and between mean or summed transverse momenta in one $\eta$ interval and the multiplicity of charged particles in another $\eta$ interval $(p_{T} - n)$. In our present analysis, we have explored first two types of FB correlation in detail. The FB correlation strength is measured in a coordinate system with origin $\eta=0$ which is the collision vertex. Two $\eta$ intervals are selected: one in the forward ($\eta>0$) and another in the backward hemispheres ($\eta<0$) in the center-of-mass system.
FB multiplicity correlation coefficient $b_{corr}(mult)$ and FB momentum correlation coefficient $b_{corr}\left(\Sigma p_{T}\right)$ have been estimated using the following formula of Pearson Correlation Coefficients~\cite{ref7}
%
\begin{equation}
\label{eq1}
b_{corr} (mult) = \frac{\langle N_{f} N_{b}\rangle - \langle N_{f}\rangle \langle N_{b}\rangle}{\langle N_{f}^{2}\rangle - \langle N_{f}\rangle^{2}}\text{,}
\end{equation}
\begin{equation}
\label{eq2}
b_{corr}(\Sigma p_{T})=\frac{\langle\Sigma p_{T_{f}}\Sigma p_{T_{b}}\rangle - \langle\Sigma p_{T_{f}}\rangle\langle\Sigma p_{T_{b}}\rangle}{\langle(\Sigma p_{T_{f}})^{2}\rangle - \langle\Sigma p_{T_{f}}\rangle^{2}}
\end{equation}
%
Here, $N_{f}$ and $N_{b}$ are the charged particle multiplicity in F and B window respectively. $\Sigma p_{T_{f}}$ and $\Sigma p_{T_{b}}$ are the event summed transverse momentum in F and B window respectively.
\section{The EPOS3 Model}
\label{sec:epos3}
The EPOS3 model is based on Gribov-Regge multiple scattering theory~\cite{ref1}. In this approach an individual scattering is labeled as a Pomeron which creates a parton ladder carrying the transverse momentum from the initial hard scatterings~\cite{ref2}. In a collision, many elementary parton-parton hard scatterings form a large number of flux tubes that expand and are fragmented into string segments. Higher string density forms ``core" which undergoes full 3D+1 viscous hydrodynamical collective expansion expecting no jet parton escape. Another part of lower string density forms ``corona" where we can expect the escape of jet partons. 
Using EPOS3 model, we have generated 3 million minimum-bias $pp$ events at $\sqrt{s}=13$ TeV, for each of the options, with and without hydro.
\section{Results}
\label{sec:FBSelec}
The whole analysis have been carried out following ALICE \cite{ref8} and ATLAS \cite{ref9} kinematics.
We have selected events having a minimum of two charged particles in the chosen kinematic intervals.
\begin{table}[h!]
\center
\begin{tabular}{|c||c|}
\hline
ALICE kinematics & ATLAS kinematics \\
\hline
0.3 < $p_{T}$ < 1.5 (GeV) and |$\eta$| < 0.8 & $p_{T}$> 0.1 (GeV) and |$\eta$| < 2.5 \\
\hline
\end{tabular}
\end{table}

\subsection{Dependence on the gap between FB windows ($\eta_{gap}$)}
\begin{figure}[!ht]
\centering
\begin{subfigure}{.5\textwidth}
\centering
\includegraphics[width=1.0\linewidth]{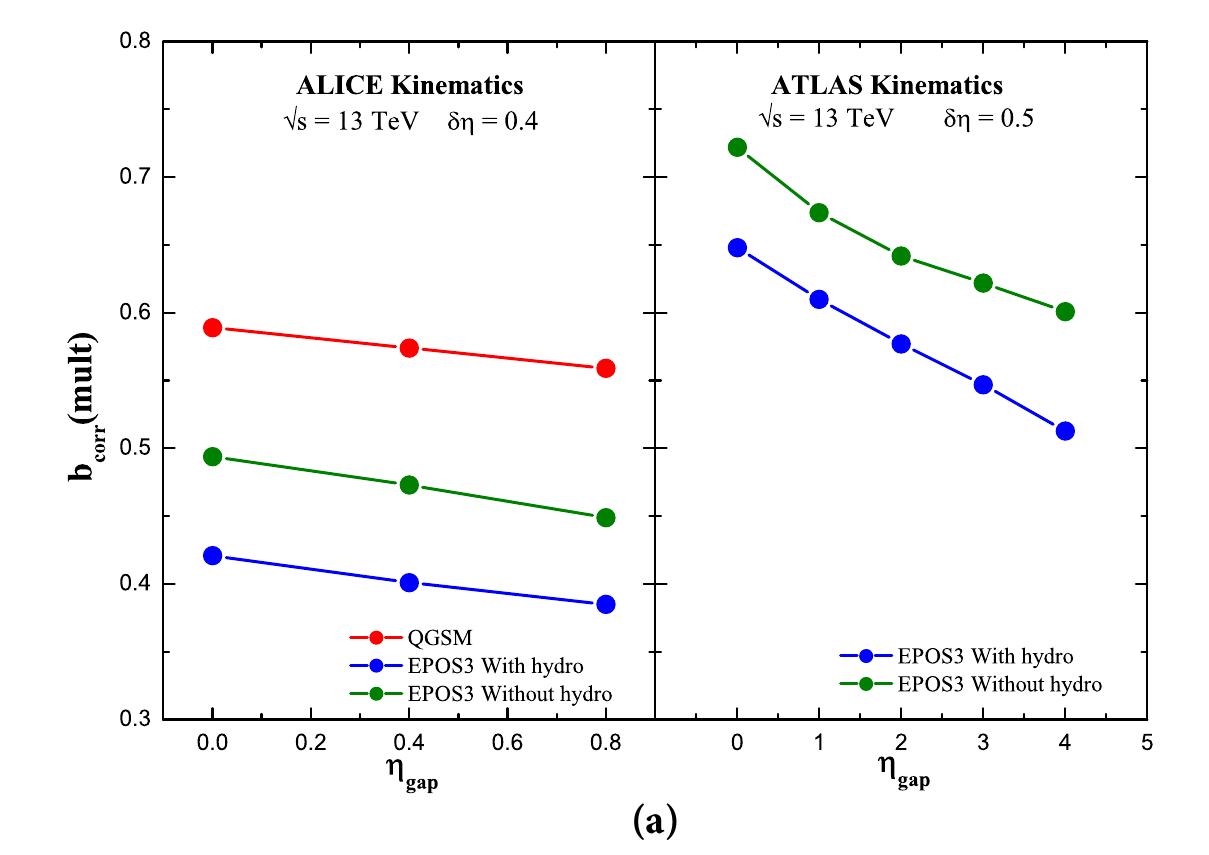}
\label{fig:sub1}
\end{subfigure}%
\begin{subfigure}{.5\textwidth}
\centering
\includegraphics[width=1.0\linewidth]{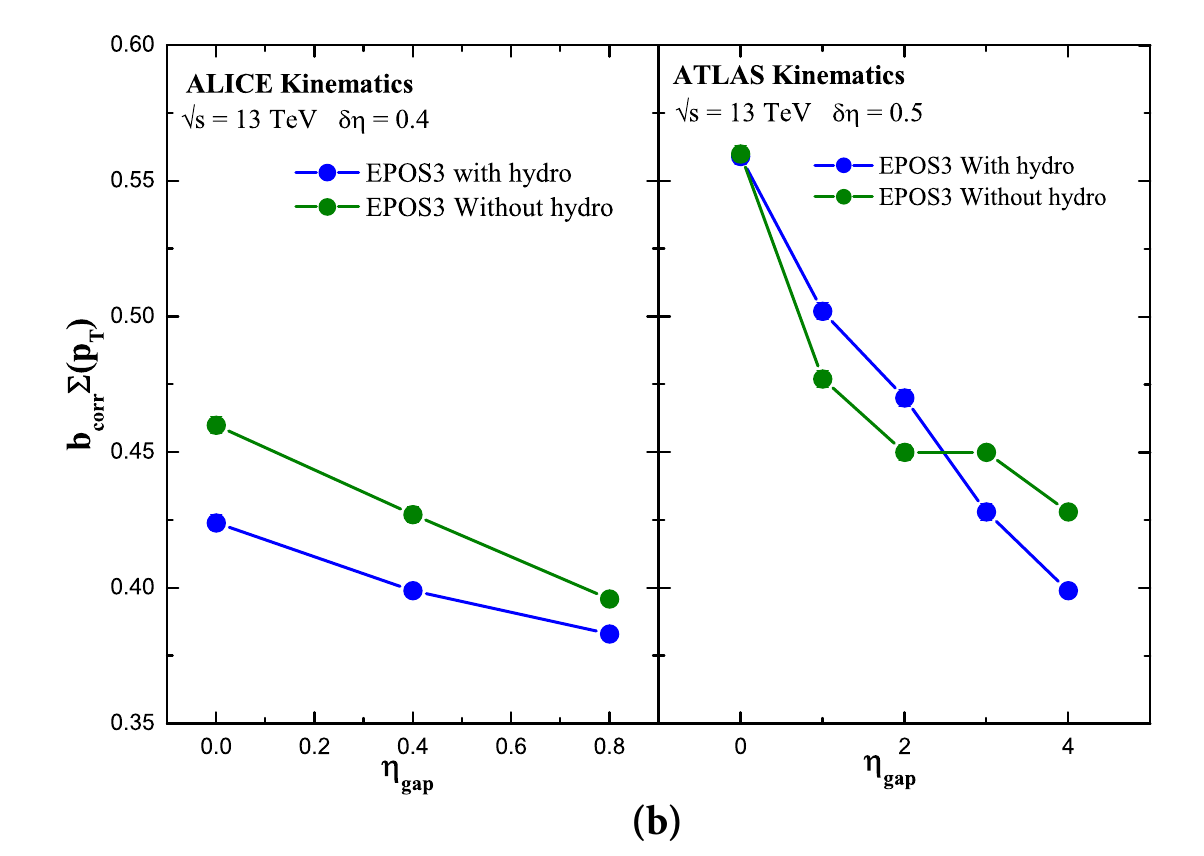}
\label{fig:sub2}
\end{subfigure}
\caption{(a) $b_{corr}(mult)$ and (b) $b_{corr}\left(\Sigma p_{T}\right)$ as a function of $\eta_{gap}$.}
\label{fig:text}
\end{figure}

The variations of FB multiplicity and momentum correlation coefficients estimated using Eq.~\ref{eq1} \& \ref{eq2} with $\eta_{gap}$ of window widths 0.4 and 0.5 for EPOS3 simulated $pp$ events with and without hydro at $\sqrt{s}=13$ TeV following ALICE (left panel) and ATLAS (right panel) kinematics have been shown in Fig.~\ref{fig:text}. FB multiplicity correlations are compared to the QGSM calculation~\cite{ref4}.
%
\subsection{Dependence on Collision Energy ($\sqrt{s}$)}
\begin{figure}[!ht]
\centering
\begin{subfigure}{0.5\textwidth}
  \centering
  \includegraphics[width=1.0\linewidth]{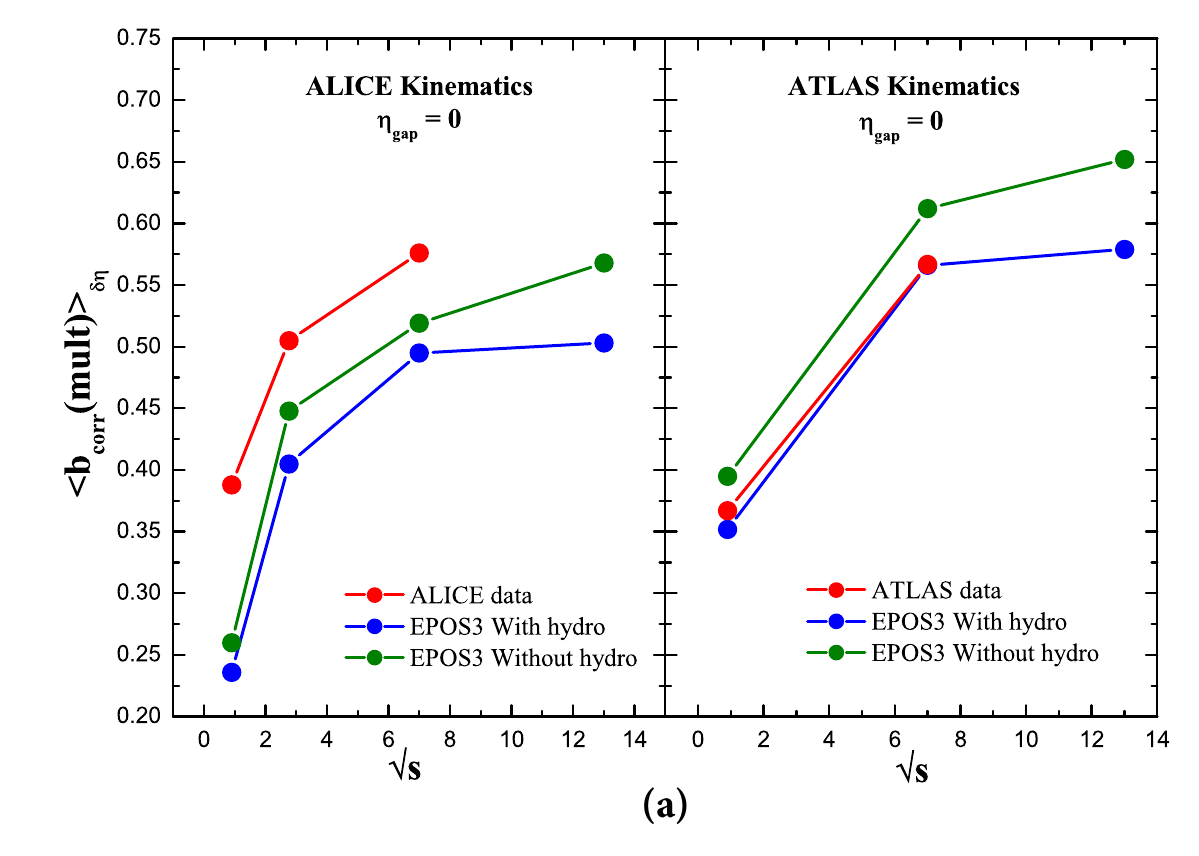}
  \label{fig:sub1}
\end{subfigure}%
\begin{subfigure}{0.5\textwidth}
  \centering
  \includegraphics[width=1.0\linewidth]{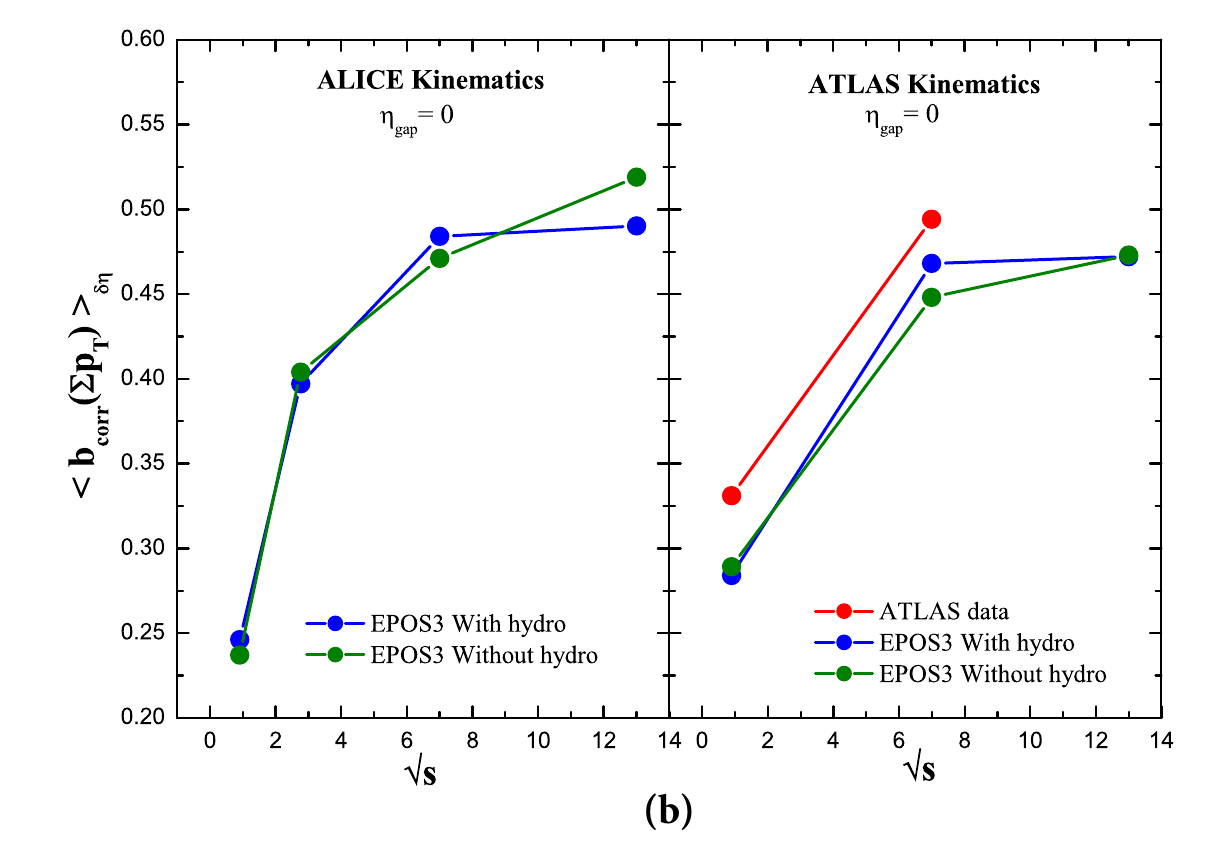}
  \label{fig:sub2}
\end{subfigure}
\caption{(a) $\left<b_{corr}(mult)\right>_{\delta\eta}$ and (b) $\left<b_{corr}\left(\Sigma p_{T}\right)\right>_{\delta\eta}$ as a function of centre-of-mass energy ($\sqrt{s}$). 
}
\label{fig:test}
\end{figure}

Center-of-mass energy dependence of $\delta\eta$-weightage average of FB multiplicity and momentum correlation is exhibited in Fig.~\ref{fig:test} for both ALICE and ATLAS kinematics. Red points represent corresponding experimental data. 
\section{Conclusion}
We have investigated the dependence of FB correlation strength on $\eta_{gap}$ at $\sqrt{s}=13$ TeV for EPOS3 simulated with and without hydro $pp$ events. 
From the study we find that the FB correlation strengths decrease slowly with the increase of $\eta_{gap}$ at $\sqrt{s}=13$ TeV for both EPOS3 hydro and without hydro events which resemble our previous observations at lower center-of-mass energies~\cite{ref3}. It is evident that our results are in qualitative agreement with the available QGSM predictions~\cite{ref4}.
We also find that the $\delta\eta$-weightage average of FB correlation strength tends to saturate at very high energy. 
\section*{Acknowledgements}
The authors are thankful to the members of the grid computing team of VECC, Kolkata and cluster computing team of Department of Physics, Jadavpur University for providing uninterrupted facility for event generation and analyses. We also gratefully acknowledge the financial help from the DHESTBT, WB. J. M acknowledges DST-INDIA INSPIRE fellowship Scheme and S. K acknowledges the financial support from UGC-INDIA DSK-PDF under Grant No. F.4-2/ 2006(BSR)/PH/19-20/0039.


\bibliography{mybib.bib}

\nolinenumbers

\end{document}